\documentclass[aps,prb,twocolumn,superscriptaddress,showpacs]{revtex4-1}
\usepackage{graphicx}
\usepackage{calc}
\usepackage{bm}

\begin{document}

\title{Andreev reflection at the edge of a two-dimensional electron system with strong spin-orbit coupling}

\author{A.~Kononov}
\affiliation{Institute of Solid State Physics RAS, 142432 Chernogolovka, Russia}
\affiliation{Moscow Institute of Physics and Technology, Institutsky per. 9, Dolgoprudny, 141700 Russia}
\author{G.~Biasiol}
\affiliation{IOM CNR, Laboratorio TASC, 34149 Trieste, Italy}
\author{L.~Sorba}
\affiliation{NEST, Istituto Nanoscienze-CNR and Scuola Normale Superiore, 56127 Pisa, Italy}
\author{E.V.~Deviatov}
\affiliation{Institute of Solid State Physics RAS, 142432 Chernogolovka, Russia}
\affiliation{Moscow Institute of Physics and Technology, Institutsky per. 9, Dolgoprudny, 141700 Russia}

\date{\today}

\begin{abstract}
We experimentally investigate transport properties of a single planar junction between the niobium superconductor and the edge of a two-dimensional electron system in a narrow $In_{0.75}Ga_{0.25}As$ quantum well with strong Rashba-type spin-orbit coupling. We experimentally demonstrate suppression of Andreev reflection at low biases at ultra low temperatures. From the analysis of temperature and magnetic field behavior, we interpret the observed suppression as a result of a spin-orbit coupling. There is also an experimental sign of the topological superconductivity realization in the present structure. 
\end{abstract}


\maketitle


Recent interest to transport investigations of hybrid superconductor-normal structures is mostly stimulated by the search for Majorana fermions, which are their own anti-particles~\cite{Wilczek}. The key feature of hybrid structures~\cite{reviews} is the allowed regime of topological superconductivity~\cite{Fu,Sau1,Potter}.

The regime of topological superconductivity is connected~\cite{Fu,Sau1,Potter} with spectrum modification in normal metal, induced by cooperation of strong spin-orbit (SO) splitting $\Delta_{SO}$, Zeeman splitting $E_Z$ and the gap $\Delta_{ind}$ induced due to a proximity with s-wave superconductor.  For moderate values of magnetic fields  ($\Delta_{SO}>E_Z$), topological superconductivity appears~\cite{Fu,Sau1,Potter}  in the regime $E_Z>\Delta_{ind}$ at low temperatures $T<<\Delta_{ind}$.  

It is clear, that experimental realization of this regime requires a normal conductor with strong SO coupling. Different types of normal conductors have been proposed: topological insulator~\cite{Fu}; two-dimensional (2D) electron systems~\cite{alicea,sarma,Nakosai}; one-dimensional (1D) semiconductor wires~\cite{Lutchyn,Oreg,Pientka}. The experimental signatures of Majorana fermions have been only obtained in the case of 1D wires~\cite{Heiblum,Mourik,Deng}. 

On the other hand, the edge of a 2D electron system is well-known to exhibit 1D behavior both in quantizing~\cite{buttiker,shklovskii} and in zero~\cite{molenkamp,kvon} magnetic fields. Thus, it is quite reasonable~\cite{Nakosai} to study charge transport in a hybrid planar device, where the normal side is a 2D structure edge. Even disregarding Majorana problem, the physics is expected~\cite{inoue} to be quite sophisticated in the case of strong SO coupling in 2D system.  

Here, we experimentally investigate transport properties of a single planar junction between the niobium superconductor and the edge of a two-dimensional electron system in a narrow $In_{0.75}Ga_{0.25}As$ quantum well with strong Rashba-type spin-orbit coupling. We experimentally demonstrate suppression of Andreev reflection at low biases at ultra low temperatures. From the analysis of temperature and magnetic field behavior, we interpret the observed suppression as a result of a spin-orbit coupling. There is also an experimental sign of the topological superconductivity realization in the present structure.


Our samples are grown by solid source molecular beam epitaxy on semi-insulating GaAs  substrates. The active layer is composed of a 20-nm thick $In_{0.75}Ga_{0.25}As$ quantum well sandwiched between a lower 50-nm thick and an upper 120-nm thick $In_{0.75}Al_{0.25}As$ barriers. Details on the growth parameters can be found elsewhere~\cite{biasiol05}. 
A two dimensional electron gas (2DEG), confined in a narrow asymmetric $In_{0.75}Ga_{0.25}As$ quantum well, is characterized by: (i) high mobility at relatively low electron concentration,~\cite{biasiol08} because the well is nominally undoped;~\cite{biasiol05} (ii) strong Rashba-type SO coupling~\cite{holmes,inas}; (iii) high $g$-factor~\cite{biasiol04}. For our samples, the 2DEG mobility at 4K is about $5 \cdot 10^{5}  $cm$^{2}$/Vs  and the carrier density is   $4.1 \cdot 10^{11}  $cm$^{-2}$, as obtained from standard magnetoresistance measurements.

\begin{figure}
\centerline{\includegraphics*[width=0.7\columnwidth]{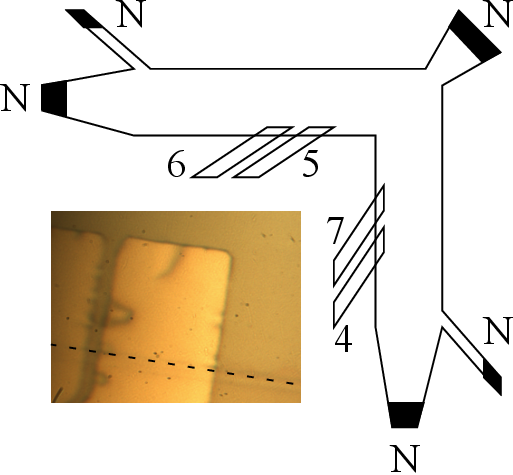}}%
\caption{(Color online) Schematic diagram of the sample. The mesa is a $\Gamma$-shape Hall-bar with a number of Ni-Au Ohmic contacts (black, denoted by N). 100 nm thick niobium stripes (denoted by numbers) are placed to overlap with mesa edges. In the overlap region,  SIN junction is formed between the Nb electrode and the 2DEG edge. An image of a single overlap region is shown in the inset. The spacing between two Nb electrodes is equal to  2~$\mu m$ along the mesa edge (highlighted by dashed line).  
\label{sample}}
\end{figure}

Schematic diagram of the sample is presented in Fig.~\ref{sample}. The mesa is a $\Gamma$-shape, 100~$\mu$m width Hall-bar with a number of Ni-Au Ohmic contacts (black). Two perpendicular mesa edges are oriented along $[011]$ and $[01\overline{1}]$ crystallographic directions. The 200 nm height mesa step is formed by wet chemical etching. 

It is important, that in our $In_{0.75}Ga_{0.25}As$ structure a high quality contact to a 2DEG edge can be realized~\cite{biasiol08} by evaporation of a metal over the mesa edge, without annealing procedure.  We thermally evaporate 10~nm Ni and 100~nm Au to obtain (normal) Ohmic contacts, see Fig.~\ref{sample}. Any normal contact is characterized by a strictly linear $I-V$ dependence at low temperatures with $\approx 500\Omega$ resistance. 
In addition, we use dc sputtering to place 100~nm thick niobium stripes to overlap with  mesa edges, see Fig.~\ref{sample}. The width of a single stripe is 20~$\mu$m in the overlap region. The stripes are formed by lift-off technique, the surface is mildly cleaned in Ar plasma before sputtering. 

The clean junction between the Nb electrode and the edge of a 2DEG is essentially a planar superconductor (S) -- insulator (I) -- normal (N) junction. Because of the edge electrostatics~\cite{shklovskii}, an electron concentration of a 2DEG is gradually diminishing by approaching the etched mesa edge, so there is a depletion region (I) of finite width at the edge~\cite{shklovskii}. This region is too narrow to affect Ohmic behavior of normal contacts. However, being placed between S and N electrodes, it defines the transport properties of SIN junction~\cite{tinkham}. It was demonstrated in Ref.~\onlinecite{batov} that this depletion region can be removed by annealing Au at the mesa edge. In the present paper we, instead, concentrate on a clean junction which is especially reasonable for a high quality 2D edge~\cite{inas} in our structures.

All measurements are performed in a temperature range 30~mK-1.2~K. Similar results are obtained from two different samples in several cooling cycles. To avoid orbital effects in a 2DEG, we use in-plane oriented magnetic field. The results are independent of the magnetic field direction within the 2D plane.

\begin{figure}
\includegraphics*[width=\columnwidth]{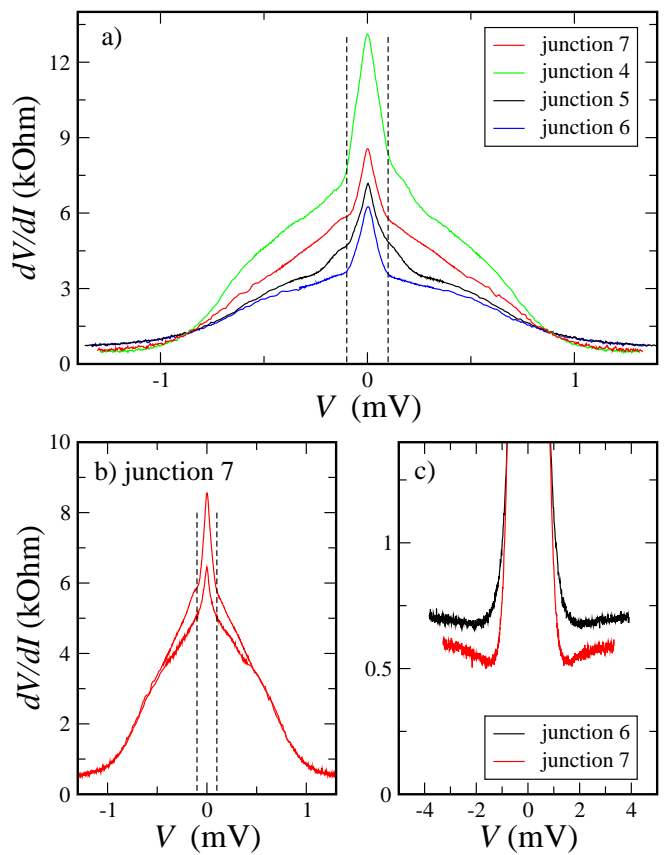}%
\caption{(Color online) (a) Differential resistance $dV/dI$ of a single SIN junction as a function of the voltage drop across the junction. The curves are denoted by the junction numbers, see Fig.~\protect\ref{sample}. The curves are characterized by a finite conductance for voltages within the superconducting gap in Nb. Each curve demonstrates the well developed $dV/dI$ resistance peak at low biases. The peak width is  denoted by dashed lines. It is the same for all the junctions. (b) The $dV/dI-V$ curves for the junction 7, obtained for two cooling cycles. The $dV/dI$ peak width is invariant to the cooling cycle. (c)  Demonstration of different normal resistance values for the junctions placed at  two perpendicular mesa edges.  
\label{IV}}
\end{figure}
\begin{figure}
\includegraphics*[width=\columnwidth]{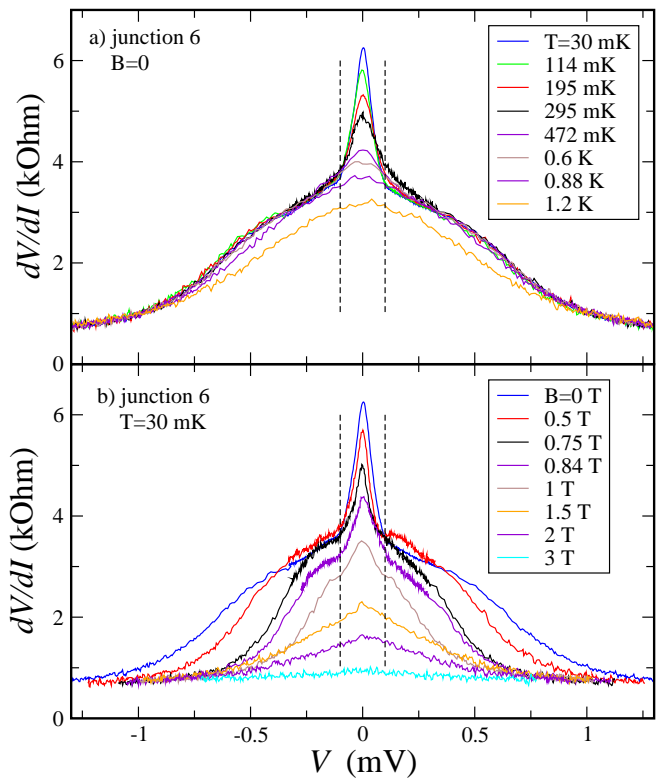}%
\caption{(Color online) The evolution of  the $dV/dI$ curve is shown with temperature (a) or in-plane magnetic field (b). The $dV/dI$  peak completely disappears at 0.88~K, while the curve is practically insensitive to the temperature below this value. The peak is fully suppressed by the magnetic field at 1.5~T, while the superconductivity survives up to $\approx 3$~T. 
\label{IVev}}
\end{figure}
\begin{figure}
\includegraphics*[width=\columnwidth]{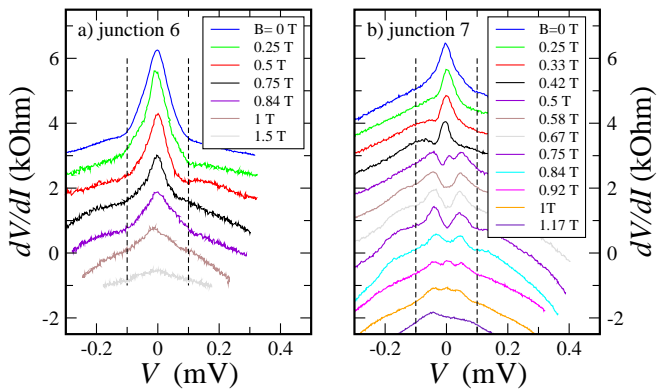}%
\caption{(Color online) Comparison of the $dV/dI-V$ curve evolution with in-plane magnetic field for two SIN junctions 6 and 7 at low biases. By contrast to the monotonous suppression for the junction 6, the junction 7, which is characterized by maximal induced gap $\Delta_{ind}$ (see the text), demonstrates sophisticated  behavior. A strong  resistance dip evolves at zero bias for magnetic fields from $B=0.5$~T. The dip is of maximum value for $B=0.58-0.75$~T and disappears above $1$~T. The curves are shifted for clarity.
\label{IV67}}
\end{figure}

We study electron transport across a single SIN junction in a three-point configuration: one Nb electrode in Fig.~\ref{sample} is grounded, two different Ohmic contacts are in use to apply a current and to measure a voltage drop across the junction. To obtain $dV/dI-V$ characteristics depicted in Figs.~\ref{IV},\ref{IVev},\ref{IV67}, we sweep dc current through the SIN junction from -1~$\mu$A to +1~$\mu$A.  This dc current is modulated by a low (0.85~nA) ac component. We measure both dc ($V$) and ac ($\sim dV/dI$) components of the voltage drop across the junction by using the electrometer and the lock-in, respectively. The resulting $dV/dI-V$ curve is independent of a particular choice of Ohmic contacts. The setup is also verified by obtaining a strictly linear $dV/dI-V$ curve if the normal Ohmic contact is grounded instead of a niobium electrode. 

The examples of  $dV/dI-V$ characteristics are presented in Fig.~\ref{IV} for different SIN junctions (a) and for different cooling cycles (b). All the curves demonstrate a finite conductance for voltages within the superconducting gap $\Delta_{Nb}$ in niobium ($T_c=9.5$~K). The conductance is partially suppressed within a narrow $dV/dI$ resistance peak at low biases. The peak width $\approx 0.2$~mV is invariant for all the junctions, see Fig.~\ref{IV} (a), and for every cooling, see Fig.~\ref{IV} (b). The curves are perfectly symmetric in respect to the voltage sign.  

At higher voltages $dV/dI$ approaches the normal resistance values, see (a) and (c), which is the same for the junctions placed at the same mesa edge, see (a). On the other hand, the normal resistance values are clearly different for two perpendicular $[011]$ and $[01\overline{1}]$ edges, which reflects intrinsic in-plane mobility anisotropy of 2DEG in $In_{0.75}Ga_{0.25}As$ structure~\cite{biasiol08}.

Fig.~\ref{IVev} demonstrates typical evolution of  the $dV/dI-V$ curve  with temperature (a) or in-plane magnetic field (b). The $dV/dI$ peak only exists at low temperatures, see panel (a). It completely disappears at 0.88~K. By contrast, the curve itself is practically insensitive to the temperature below 0.88~K,  because of much higher $T_c=9.5$~K in niobium. 

In-plane magnetic field $B$ suppresses the nonlinearity of the $dV/dI-V$ curve, see Fig.~\ref{IVev} (b). The $dV/dI-V$ curve is only slightly non-linear at $B=3$~T. The $dV/dI$ resistance peak is even more sensitive to the in-plane magnetic field. The $dV/dI-V$ curve to both sides of the peak region is not affected by the field up to $0.75$~T, while the peak amplitude is twice smaller at this field. The central  peak  disappears completely at $\approx 1.5$~T. 

We check by independent measurement, that the  niobium electrodes are in a superconducting state below  $B=3.8$~T. Resistance at high voltages $eV>\Delta_{Nb}$ is insensitive to the magnetic field, so the field is well oriented to the 2D plane.  For the measurements in zero field, see Figs.~\ref{IV} and \ref{IVev} (b), the magnetic field is well compensated which we check by reverting the solenoid polarity.


To start the discussion, let us neglect for a while the central $dV/dI$ peak at low biases, e.g. by considering a  $dV/dI-V$ curve above 0.88~K. In this case the curve demonstrates the standard behavior of the SIN junction, which can be well understood within the framework of BTK theory~\cite{tinkham}. 

In our experiment, see Figs.~\ref{IV},\ref{IVev}, the differential resistance $dV/dI$ at  $eV<\Delta_{Nb}$ is increased with respect to the normal resistance of the junction, so the potential barrier (depletion region) is present at the SN interface~\cite{tinkham}. On the other hand, the  conductivity is still finite at  $eV<\Delta_{Nb}$. The finiteness is demonstrated  not only by low absolute values of the junction resistance (20~$\mu$m junction is wide), but mostly by the clear visible suppression of the conductivity within the $dV/dI$ resistance peak at low biases. Finite conductance at voltages $eV<\Delta_{Nb}$ is only allowed because of Andreev reflection, since a single-particle tunneling is prohibited within the superconducting gap~\cite{tinkham}.   

The curves in Fig.~\ref{IV} are qualitatively similar. The difference in absolute values reflects the fact that the depletion region width varies for different junctions (a) and for different coolings (b). Andreev reflection is extremely sensitive to the potential barrier strength (i.e. depletion region width), because the Andreev process requires two-particle co-tunneling.~\cite{tinkham}

It's worth mentioning, that the normal resistance of the junction is independent of the cooling. In the regime of normal transport the junction is not sensitive to a depletion region, which is demonstrated by strictly linear $dV/dI-V$ curves in high magnetic fields and by the Ohmic behavior of normal contacts. The same normal resistance at a single mesa edge indicates roughly similar short-range disorder (which defines the mobility) along the particular edge. We can estimate a single-particle transmission $Т$ to be about 0.3 from the value of the junction normal resistance and the junction width.

Let us highlight the most important experimental result: a strong increase of the resistance within 0.2~mV interval at low biases indicates a suppression of the Andreev reflection, and this behavior is universal for different junctions, samples, and coolings. 

This suppression is very unusual for SN structures with a 2DEG, see, e.g., Ref.~\onlinecite{beltram}. It can not result from the residual Schottky barrier at a 2DEG edge: (i) there is no noticeable Schottky barrier at the edge of the  $In_{0.75}Ga_{0.25}As$ structure~\cite{biasiol08}; (ii) the $dV/dI$ resistance peak is perfectly symmetric with respect to the voltage sign; (iii) $dV/dI-V$ curves are strictly linear for normal Ohmic contacts and for S-type contacts in high magnetic fields.

We have to connect the observed suppression of the  with a strong Rashba-type spin-orbit  coupling in our 2DEG.~\cite{inas} Rashba-type SO coupling mixes spin-up and spin-down states. It induces an energy splitting $\Delta_{SO}$ which lifts the spin degeneracy, but the energy splitting does not break the time reversal symmetry unlike an exchange splitting in ferromagnet. The calculation in Ref.~\onlinecite{inoue} indicates, that the Andreev reflection  is indeed suppressed in the case of the insulating barrier of intermediate strength and strong SO coupling ($Z=1$ and $\beta=0.2 - 0.6$ in terms of Ref.~\onlinecite{inoue}). 

If the voltage $V$ or temperature $T$ exceeds $\Delta_{SO}$, the spin degeneracy is restored, and therefore the Andreev reflection. For our planar junctions it is crucial that $\Delta_{SO}\sim \alpha k_F$ is seriously reduced~\cite{holmes} at the 2DEG edge, because of reduced electron concentration~\cite{shklovskii}. 

We can expect the minimal edge concentration of delocalized electrons to be less than $1 \cdot 10^{11}  $cm$^{-2}$, in agreement with  different independent  measurements~\cite{biasiol08,holmes,inas} of gated $In_{0.75}Ga_{0.25}As$ structures. In this case, the Rashba coupling constant $\alpha$ can be expected~\cite{holmes} to be below $0.5 \cdot 10^{-11}$~eVm. It is worth noting that even this value exceeds a typical bulk $\alpha$ for AlGaAs structures by an order of magnitude. 

If we assume that the central peak width $\Delta V \approx 0.2$~mV is defined by $\Delta_{SO}$ at the edge, we can estimate $k_F=\Delta_{SO}/\alpha \approx 4 \cdot 10^{5}  $cm$^{-1}$ at the edge. This corresponds to a minimal electron concentration $\approx 0.3 \cdot 10^{11}  $cm$^{-2}$ at the edge, in a reasonable agreement with values obtained in Ref.~\onlinecite{holmes} for dilute $In_{0.75}Ga_{0.25}As$ structure. This estimation $\Delta_{SO}/2 \approx 0.1\mbox{ meV}$ is also consistent with the resistance peak disappearance at $T\approx 1$~K.  We can therefore use the field $B=1.5$~T of the resistance peak suppression as a crude estimation~\cite{alicea} of $\Delta_{SO}\sim E_Z$. This estimation is  in a reasonable agreement with values obtained in Ref.~\onlinecite{holmes}.

From the above description it is clear that the present SIN junction is a good candidate to realize the regime of the topological superconductivity~\cite{Fu,Sau1,Potter,alicea}. Indeed, in the magnetic fields below 1.5~T the spin-orbit coupling exceeds~\cite{holmes} the Zeeman splitting $\Delta_{SO}>E_Z$. On the other hand, niobium electrode induces superconductivity in the neighbor 2DEG at the sample edge. The induced gap $\Delta_{ind}$ strongly depends on the width of the depletion region~\cite{tinkham} and  is always smaller than the gap in Nb: $\Delta_{ind}<\Delta_{Nb}$. The induced gap $\Delta_{ind}$ is diminishing to the bulk of the sample, so the topological superconductivity regime  $\Delta_{ind}<E_Z$ is always realized within some region at the edge (the helical Majorana edge channels in Ref.~\onlinecite{Nakosai}). 

The signature of this behavior might be seen in Fig.~\ref{IV67} for the junction 7. Fig.~\ref{IV67} (b) demonstrates that  a strong and narrow resistance dip (i.e. conductance peak) evolves at zero bias for magnetic fields higher than $B=0.5$~T and disappears above $B=1$~T. This behavior is observed for both coolings of the sample. The resistance dip can not be ascribed to the single-particle potential barrier~\cite{inoue,tinkham} within the depletion region, since it is only present for a narrow magnetic field range. 

It is important, that this narrow dip is observed for the parameter range which well satisfy the conditions of the topological superconductivity~\cite{Fu,Sau1,Potter,alicea}.  We can expect a reasonable $\Delta_{ind}>T$ for this junction 7: (i) it is placed at the mesa edge which is of minimal  disorder; (ii)  it is of a minimal  differential resistance at this mesa edge, so it is characterized by smallest depletion region and, therefore, by maximum induced gap $\Delta_{ind}$ in 2DEG.  The narrow resistance dip in Fig.~\ref{IV67} (b) is centered within the wider spin-orbit resistance peak (denoted by dashed lines in Fig.~\ref{IV67}), so $\Delta_{SO}>E_Z$ is fulfilled. At higher fields this relation is broken, which is accompanied by disappearance of the resistance dip above 1~T. Thus, the transport behavior of the junction 7 is in agreement with that one can expect for the topological superconductivity regime at the 2DEG edge~\cite{Nakosai}.



We wish to thank  Ya.~Fominov for fruitful discussions and S.V.~Egorov for help in dc sputtering.
We gratefully acknowledge financial support by the RFBR (project No. 13-02-00065) and RAS.

\end{document}